# Navigating AI to Unpack Youth Privacy Concerns: An In-Depth Exploration and Systematic Review

Ajay Kumar Shrestha
*Computer Science Department*
*Vancouver Island University*
Nanaimo, Canada
ajay.shrestha@viu.ca

Ankur Barthwal
*Business and Management*
*Vancouver Island University*
Nanaimo, Canada
ankur.barthwal@viu.ca

Molly Campbell
*Computer Science Department*
*Vancouver Island University*
Nanaimo, Canada
molly.campbell@viu.ca

Austin Shouli
*Computer Science Department*
*Vancouver Island University*
Nanaimo, Canada
austin.shouli@viu.ca

Saad Syed
*VIU Affiliate*
*Vancouver Island University*
Nanaimo, Canada
saad.syed@viu.ca

Sandhya Joshi
*VIU Affiliate*
*Vancouver Island University*
Nanaimo, Canada
sandhya.joshi@viu.ca

Julita Vassileva
*Computer Science Department*
*University of Saskatchewan*
Saskatoon, Canada
jiv@cs.usask.ca

***Abstract*— This systematic literature review investigates perceptions, concerns, and expectations of young digital citizens regarding privacy in artificial intelligence (AI) systems, focusing on social media platforms, educational technology, gaming systems, and recommendation algorithms. Using a rigorous methodology, the review started with 2,000 papers, narrowed down to 552 after initial screening, and finally refined to 108 for detailed analysis. Data extraction focused on privacy concerns, data-sharing practices, the balance between privacy and utility, trust factors in AI, transparency expectations, and strategies to enhance user control over personal data. Findings reveal significant privacy concerns among young users, including a perceived lack of control over personal information, potential misuse of data by AI, and fears of data breaches and unauthorized access. These issues are worsened by unclear data collection practices and insufficient transparency in AI applications. The intention to share data is closely associated with perceived benefits and data protection assurances. The study also highlights the role of parental mediation and the need for comprehensive education on data privacy. Balancing privacy and utility in AI applications is crucial, as young digital citizens value personalized services but remain wary of privacy risks. Trust in AI is significantly influenced by transparency, reliability, predictable behavior, and clear communication about data usage. Strategies to improve user control over personal data include access to and correction of data, clear consent mechanisms, and robust data protection assurances. The review identifies research gaps and suggests future directions, such as longitudinal studies, multicultural comparisons, and the development of ethical AI frameworks. The findings have significant implications for policy development and educational initiatives aimed at empowering young users to manage their data effectively, addressing the evolving privacy landscape shaped by AI technologies.**

## I. INTRODUCTION

Today's youth, referred to as young digital citizens, defined as children and young people raised in a technology-driven world, frequently encounter challenges related to the misuse and unauthorized access of their personal data [1]. High-profile data breaches and misuse of data on social media have heightened these concerns [2], [3], [4]. Many young digital citizens, particularly those aged 16-19, use AI-driven applications but often lack understanding of how these systems work, making them more vulnerable to privacy risks. This age group is particularly significant as they are at a critical stage of gaining digital independence and forming key attitudes toward privacy and data sharing. Their data-sharing practices are influenced by perceived benefits, strong data protection assurances, and transparency concerns. They are more likely to share data if they perceive tangible benefits and robust security measures [5], [6]. Explicit privacy policies and clear data usage practices are crucial to build trust.

Given these challenges and the complex landscape of privacy in AI systems, a systematic literature review is essential to comprehensively understand the perceptions, concerns, and expectations of young digital citizens regarding their privacy. This systematic literature review synthesizes findings from various studies to answer key questions about privacy concerns, data-sharing practices, balancing privacy and utility, trust factors, transparency expectations, and strategies for enhancing user control over personal data, with a particular focus on young users.

While young digital citizens value the personalized services and enhanced experiences offered by AI, they remain wary of potential privacy risks. This systematic review highlights the need for transparent data-sharing practices, ethical AI frameworks, and ongoing education to effectively address these concerns [7], [8]. Trust in AI systems among young digital citizens hinges on transparency, reliability, and ethical standards, making clear communication about data practices, consistent performance, and user-friendly interfaces essential [9], [10]. Transparency expectations vary across contexts, with greater demands in healthcare and education compared to social media. Enhancing user control over personal data is also crucial, including providing access and correction rights, clear consent mechanisms, and robust data protection assurances. Young digital citizens particularly value managing their information and demand straightforward consent processes [11], [12].

The review also identifies several research gaps and future directions, such as the need for longitudinal studies to track changes in privacy attitudes over time, multicultural comparisons to understand global perspectives and the development of ethical AI frameworks that incorporate young users' views. Addressing these gaps will contribute to a comprehensive understanding of privacy in AI systems and inform the creation of user-centric, ethical AI technologies [2], [13]. These findings can have implications for policy development, educational initiatives, and the design of AI technologies that align with the privacy needs and expectations of young digital citizens. Ultimately, this research aims to provide insights into the development of AI systems that are transparent, reliable, and respectful of user

The Office of the Privacy Commissioner of Canada's (OPC) Contributions Program.

privacy, thereby contributing to a safer and more trustworthy digital environment for young users.

The remainder of the paper is structured as follows: Section II details the review process used in the study. Results are presented in Section III, with the discussion and limitations following in Section IV. Finally, Section V concludes the paper.

## II. METHODOLOGY

We used a systematic review methodology [14], [15], as illustrated in Fig. 1, to explore and synthesize existing research on the perceptions, concerns, and expectations of young digital citizens regarding privacy in AI systems. We aimed to provide a comprehensive understanding of the current knowledge, identify key themes, and pinpoint areas where further research is needed. The process included data collection, screening, and analysis to ensure a reliable review of the state-of-the-art literature. We prioritized studies with clear methodologies, representative samples, and robust analysis, while noting those with significant bias risks but still including them for context. We have included a focused selection of key references that are most critical to the main themes and findings of our paper.

### A. Research Questions

These questions provide a structured framework for exploring the various dimensions of privacy issues in AI.

- *RQ1: What are the predominant privacy concerns expressed by young digital citizens about AI technologies?*
- RQ2: How do young digital citizens perceive the balance between privacy and utility in AI-driven applications?
- RQ3: What are the current trends in research regarding youth attitudes toward data-sharing practices in AI systems?
- RQ4: What factors influence the trust of young digital citizens in AI-driven applications?
- RQ5: How do transparency expectations vary among young digital citizens concerning the use of AI technologies across different contexts?
- RQ6: What strategies have been proposed in the literature to enhance user control over personal data in AI applications, and how effective are they perceived by young digital citizens?
- RQ7: What are the identified research gaps and future directions for studying the perspectives of young digital citizens on privacy within AI systems?

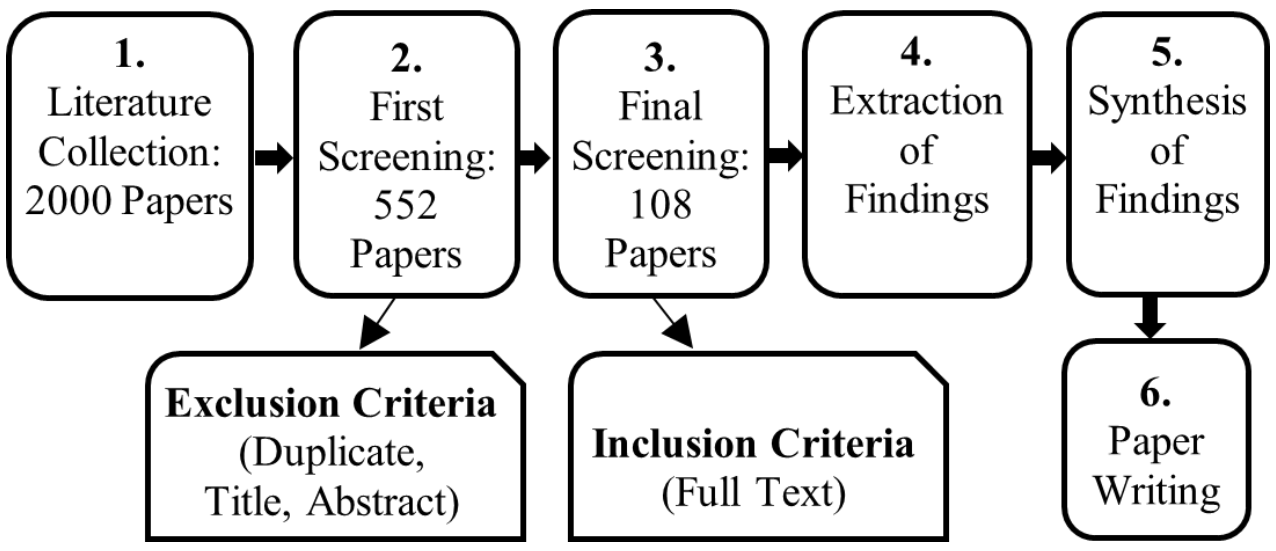

Fig. 1. Research Methodology

### B. Literature Collection

The data collection process followed a systematic approach to identify and gather relevant literature from various academic sources, ensuring comprehensive coverage of studies related to the privacy perceptions of young digital citizens in the context of AI systems. To ensure a broad and thorough coverage of relevant studies the literature search used databases that publish papers in the area of AI applications: Google Scholar, IEEE Xplore, ScienceDirect, Scopus, ACM Digital Library, and Springer. To ensure comprehensive coverage, we applied a range of keywords and search terms including "AI and data privacy expectations," "AI and digital natives," "Data sharing concerns in AI," "Generative AI and Privacy," "Privacy attitudes in AI systems," "Privacy perceptions in AI," "Transparency expectations in AI," "User control over data," "Youth perceptions of AI technologies," "Youth perspectives on AI privacy," and "Young digital citizens." These keywords were combined using Boolean operators (AND, OR, NOT) to refine the search. Filters were applied to focus on recent publications (within the last 5-10 years), peer-reviewed articles, and those written in English. We employed Python scripts to scrape papers from databases and conducted manual searches to supplement the automated search results. The Python scripts used for this process are available in our GitHub project repository [16].

### C. Screening and Selection

This stage involved systematically screening and selecting studies to ensure the inclusion of only the most relevant and high-quality papers. To maintain objectivity and consistency, two researchers independently reviewed the papers and compared their decisions on exclusion. The initial screening used Excel's "Remove Duplicates" function to eliminate duplicates, followed by a quick scan of titles and abstracts to discard irrelevant studies. Papers were labeled as "Relevant" or "Not Relevant" based on their alignment with the research questions, reducing the pool from 2,000 to 552 papers. To further refine the selection, detailed inclusion and exclusion criteria were applied. The inclusion criteria focused on papers addressing privacy concerns, data-sharing practices, transparency, user control, and trust in AI among young digital citizens. Exclusion criteria filtered out studies not focusing on AI, not considering young digital citizens, non-peer-reviewed studies, and studies not in English. Abstracts were read more thoroughly to apply these criteria, and papers were marked as "Include" or "Exclude" accordingly, with the context of each paper captured in an additional column in the Excel file. For the full-text screening, the full texts of papers marked as "Include" were obtained through academic databases, institutional access, or direct contact with authors. These full texts were reviewed to ensure they met the relevance and criteria standards. Papers passing this stage were marked as "Final Include," resulting in a final selection of 108 papers, ensuring the inclusion of only the most pertinent and high-quality research for the review.

### D. Extraction of Findings

Key findings from the included studies were carefully extracted to address each research question. This process involved documenting essential elements including study objectives, methodologies, sample sizes, demographics, and key findings. The extracted information was systematically organized in the Excel sheet to facilitate thorough analysis, ensuring a comprehensive and structured review of the relevant research.

### E. Synthesis of Findings

The extracted key findings and insights were systematically organized and synthesized to identify common themes, patterns, and trends. This process involved sorting and filtering data in Excel and noting any conflicting findings or areas that require further research. Extracted information was grouped by themes related to the research questions, with recurring patterns and significant literature gaps highlighted.

### F. Paper Writing

The final review paper writing process began with summarizing the key findings from the included studies for each research question, highlighting the most significant insights and contributions. Based on the synthesis, research gaps were identified, and future research directions were proposed. A comprehensive literature review paper was then drafted, integrating the summarized findings and discussing the identified research gaps.

## III. RESULTS

For this review, only papers that specifically addressed the experiences and concerns of young users were selected. This included both user studies and general discussions or opinion papers, as long as they focused on young digital citizens. Through thematic analysis, seven key themes were identified, including Privacy Concerns in AI, Data Sharing Practices, Privacy-Utility Equilibrium, Trust Factors in AI, Transparency Expectations, User Control Over Data, and Research Gaps and Future Directions. A pie chart in Fig. 2 illustrates the distribution of papers discussing these themes. Each research question outlined in this Section is addressed through thematic analysis, with subsequent subsections providing insights and answers to each question based on the thematic findings.

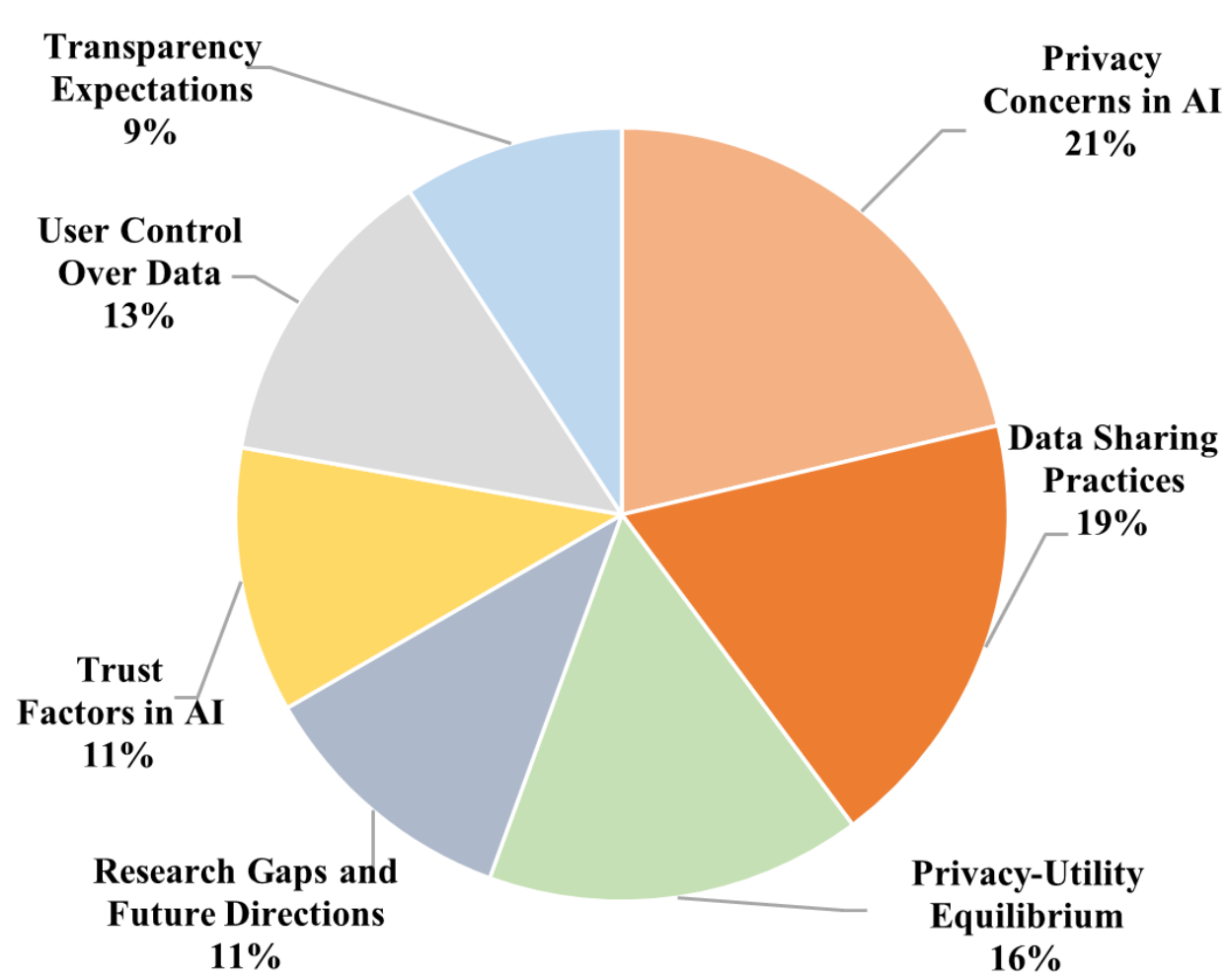

Fig. 2. Distribution of Papers Across Themes

### A. What are the predominant privacy concerns expressed by young digital citizens about AI technologies?

This research question focuses on the themes of Privacy Concerns in AI and Data Sharing Practices. Young digital citizens have significant privacy concerns about AI technologies, stemming from their interactions with digital environments. Research shows that many young users feel disempowered by AI data collection, contributing to stress and discomfort [2], [3]. This vulnerability is made worse by frequent reports of data breaches and unauthorized access to personal information, further eroding trust in these technologies [5].

The lack of transparency in data collection practices also intensifies privacy concerns [9], [10]. Many young users believe that AI systems do not sufficiently explain how their data is collected, stored, and used, leading to increased suspicion and diminished trust [9], [10]. They also worry that their personal information could be exploited for purposes beyond their control, such as targeted advertising or malicious activities [4]. Security issues, including high-profile data breaches, further raise anxiety about the vulnerability of personal data [3], [17]. This overall lack of control, transparency, and security shapes the predominant privacy concerns among young digital citizens. These concerns are further amplified by the complexity and diversity of data sources managed by AI systems, such as financial data, health records, and private communications, as highlighted in [18]. This complexity underscores the need for robust data protection measures tailored to these sensitive types of information. The same study indicates that socio-cultural factors also play a role, as young digital citizens from various regions may have differing privacy priorities. For instance, some cultures may place greater emphasis on data anonymity and minimal data collection, while others prioritize data utility and innovation. This diversity calls for localized privacy strategies to effectively address these unique needs [18]. Moreover, the potential for AI systems to perpetuate inequalities and biases is a growing concern among young users [19]. There is a significant worry about the fairness and accountability of AI, particularly in sensitive applications such as education, recruitment and social services, as noted in [20]. Addressing these concerns requires a comprehensive approach that includes technical solutions such as developing unbiased algorithms as well as ethical and regulatory measures to ensure fairness and accountability in AI systems [19].

### B. How do young digital citizens perceive the balance between privacy and utility in AI-driven applications?

This research question explores the theme of Privacy-Utility Equilibrium, a prominent theme in existing literature. Young digital citizens acknowledge the substantial benefits provided by AI technologies, such as personalized services, enhanced user experiences, and educational advancements as highlighted in [21], [22]. However, these advantages are often tempered by privacy concerns, leaving young users in a dilemma as they weigh the benefits of AI against potential risks to their personal data, reflecting broader uncertainty about engaging with these technologies [8], [23].

The need for transparent data practices is emphasized, as clear communication about how data is collected, used, and protected can help alleviate privacy concerns and build trust [9], [10]. Despite the significant utility offered by AI, privacy concerns frequently temper enthusiasm, highlighting the critical need for policies that ensure robust data protection while preserving the benefits of AI technologies. This balance is particularly evident in healthcare and education, where AI offers significant benefits but also poses substantial privacy risks [12]. In healthcare, young users are willing to share data for better services but are wary of misuse, highlighting the need for strong data security and transparency [12]. In education, AI-driven personalized learning tools raise concerns about data confidentiality among students and parents, requiring clear communication and robust privacy protections [12]. Furthermore, the impact of AI on social dynamics, such as the potential for surveillance and the erosion of personal boundaries, adds another layer of complexity to the privacy-utility balance [7]. Addressing these issues necessitates a comprehensive approach combining technical, ethical, and regulatory measures to safeguard privacy while leveraging AI's benefits [24].

### *C. What are the current trends in research regarding youth attitudes toward data-sharing practices in AI systems?*

This research question explores the theme of Data Sharing Practices, emphasizing that young digital citizens are conditionally willing to share their data when they perceive clear benefits and feel assured of data protection [6], [25]. Ethnographic research [6] indicates that young people often mistrust AI but lack an understanding of its functions and their control over it. This study, involving three projects with YR Media found that creating media about AI ethics and data-sharing practices significantly improved youth understanding and trust, especially in culturally relevant and interactive learning environments. Parental mediation also plays a crucial role, as active discussions about data privacy led to more cautious and informed data-sharing behaviors among young users [4], [18]. Corcoran et al. [2], [13], found that youth privacy concerns increase when parents actively educate them about online privacy, compared to merely discussing online activities without in-depth education. These findings suggest that educational initiatives focused on digital literacy and data privacy can improve young users' attitudes toward data sharing and highlight the importance of privacy [26], [27].

Additionally, social incentives significantly influence data-sharing practices, as young digital citizens often share personal information for social approval or exclusive content access [28]. This underscores the need for privacy education programs that address these social dynamics, helping users make informed decisions about their data [26]. The role of digital platforms is also pivotal, with those providing clear, accessible information about data usage and protection more likely to gain young users' trust [25], [29]. A study surveying younger and older adults found no significant age differences in privacy attitudes, suggesting that transparent communication and effective privacy strategies are essential for fostering trust across all age groups [3]. Furthermore, regulatory frameworks impact data-sharing behaviors, as young users in regions with stringent data protection laws tend to be more cautious about sharing personal information [30]. This calls for harmonized privacy regulations across regions to ensure consistent protection for young users' data [31].

### *D. What factors influence the trust of young digital citizens in AI-driven applications?*

This research question addresses the theme of Trust Factors in AI, highlighting that trust in AI-driven applications among young digital citizens hinges on transparency and reliability. AI systems that clearly communicate their data practices and consistently deliver accurate outputs are more likely to inspire confidence and reduce anxiety [32], [33]. Ethical considerations, such as fairness and accountability, also significantly impact trust, with young users particularly sensitive to issues of bias in AI systems [19], [34]. The inclusion of user-friendly interfaces and children's perspectives in AI design can address these concerns, as evidenced by adolescents managing their privacy on platforms like TikTok through private channels and multiple accounts to control their exposure and build trust [10], [35].

Social influence and peer recommendations also play a critical role in trust-building, as young users tend to trust AI applications endorsed by peers or influencers, highlighting the importance of social validation [4]. Additionally, a study on emotion-sensing devices showed that traditional models, like Davis's "Technological Acceptance Model," are insufficient for addressing cross-cultural privacy concerns, emphasizing the need for AI systems aligned with diverse societal values [8]. Past experiences with AI technologies also shape trust; positive interactions foster long-term trust, while negative experiences lead to skepticism [33]. Therefore, incorporating ethical guidelines and ensuring consistent, positive user experiences are crucial for building trust in AI among young digital citizens.

### *E. How do transparency expectations vary among young digital citizens concerning the use of AI technologies across different contexts?*

This research question examines the theme of Transparency Expectations in AI. Young digital citizens have varying transparency expectations based on context, demanding higher transparency in sensitive areas like healthcare and education, where they seek detailed information about data usage and access [31], [36]. In contrast, while transparency is important in social media, expectations may be lower, yet frequent data breaches have increased the demand for clearer data policies even in casual contexts [37], [38]. Menon et al. [38] found that intrinsic motivation and privacy concerns are particularly influential in smart speaker adoption among teenagers. Across all contexts, young users consistently seek straightforward, accessible information on how their data is collected, used, and protected, which is crucial for building trust and promoting positive engagement with AI technologies [39], [40].

In social media, transparency expectations are driven by frequent incidents of data misuse. Young users demand clear information on data collection, usage, and sharing to rebuild trust [10]. Platforms should adopt transparent data practices, provide detailed explanations of policies, and involve young users in developing these measures to make the platforms more trustworthy[2], [41]. In healthcare, transparency is vital due to the sensitivity of health data. Systems that show real-time access logs and allow users to see who has accessed their data can enhance trust [9], [30]. Similarly, in education, young users and educators require clear information about how AI tools use their data. Providing detailed policies and engaging students in discussions about data privacy can build trust and ensure a positive educational environment [22], [23], [42].

### *F. What strategies have been proposed in the literature to enhance user control over personal data in AI applications, and how effective are they perceived by young digital citizens?*

This research explores the theme of User Control Over Data, emphasizing the importance of intuitive controls for accessing, modifying, and deleting personal information, which are well-received by young digital citizens [12]. Clear consent processes and transparent consent mechanisms significantly enhance user engagement and trust in AI applications [11], [43]. Implementing user-friendly interfaces and privacy-enhancing technologies (PETs), such as data anonymization and encryption, can further protect personal data while maintaining AI functionality, increasing confidence in these systems [9], [44]. Feedback mechanisms that allow users to report and resolve privacy issues empower users and foster a sense of control [30]. Involving young users in the design and development of AI systems ensures their needs are met, making AI technologies more trustworthy and user-friendly [26]. Educational initiatives on AI and data privacy, such as workshops and interactive modules, help young users understand their data rights and the importance of consent, fostering autonomy in the digital landscape [22], [45], [46].

Regulatory frameworks play a crucial role in enhancing user control over personal data. Policies that mandate transparent data practices and require explicit user consent before data collection help protect young digital citizens and build trust [47], [10]. Contextual data control features, such as tracking academic progress in educational settings or viewing access logs in healthcare, further enhance user experience and data management relevance [30]. Privacy-preserving techniques like federated learning, which train AI models on decentralized data sources, reduce the risks associated with centralized data storage and improve confidence in data security [48]. Ongoing communication about data policies, security measures, and privacy features is essential for maintaining a transparent and trust-based relationship, especially with young digital users [9].

### G. What are the identified research gaps and future directions for studying the perspectives of young digital citizens on privacy within AI systems?

This research question highlights the theme of Research Gaps and Future Directions. Despite the valuable insights, significant research gaps remain, suggesting areas for further exploration. Notably, there is a lack of longitudinal studies on how privacy attitudes evolve as AI plays a larger role in daily life, necessitating long-term research [49], [50]. Additionally, more research on cultural comparisons is needed to understand global perspectives on AI and privacy, which is crucial for developing inclusive privacy policies [51], [52]. Future research should also focus on guidelines for data ownership, consent processes, and ethical use of AI technologies.

Another significant research gap is the need for more empirical studies on the practical implementation of PETs in AI systems. Investigating how technologies like differential privacy and homomorphic encryption can be integrated into AI applications is crucial for developing user-centric privacy solutions [44], [47]. Understanding the technical challenges of these integrations is essential for advancing privacy protections [9]. Additionally, research should explore the impact of emerging AI technologies, such as generative AI, federated learning with blockchain, and emotion-sensing AI, on privacy perceptions and behaviors among young digital citizens [8]. These technologies pose unique privacy challenges, such as misuse of generated content and invasive emotion-sensing capabilities [53]. Exploring how young users perceive the trade-offs between these technologies' benefits and risks can help craft targeted privacy protections and ethical guidelines [26], [54]. Moreover, improving AI interface design is critical, as many young users find current interfaces confusing, limiting their ability to manage privacy settings effectively [55]. Future research should focus on designing more intuitive interfaces to enhance user control over privacy settings [34].

Investigating the impact of educational programs on AI literacy and privacy awareness is essential. Research should focus on how different educational strategies improve young users' understanding of AI technologies and their ability to manage personal data. Effective educational initiatives can empower young digital citizens to navigate the digital landscape safely [41]. For instance, incorporating AI and data privacy topics into school curricula and offering hands-on activities like coding workshops or privacy simulations can enhance practical knowledge and confidence in data management [34], [56]. Additionally, examining the interplay between regulatory frameworks and privacy perceptions can assess the effectiveness of current data protection laws and guide the development of more adaptive regulations [47]. Analyzing enforcement and compliance challenges can help policymakers refine global data protection measures [25]. Furthermore, there is a pressing need to address the ethical implications of AI technologies, particularly concerning biases and fairness [12]. Research should focus on developing frameworks to ensure AI systems promote equity and do not reinforce existing societal biases [51]. This includes exploring methods to detect and mitigate biases in AI algorithms and assessing the societal impacts of biased systems on various demographic groups [8]. By tackling these ethical concerns, researchers can help create AI systems that are both technically robust and socially responsible, respecting the privacy rights of young digital citizens.

## IV. Discussion

This discussion synthesizes crucial insights, outlines the implications for policy and practice and highlights the limitations of this literature review.

### A. Privacy Concerns and Control

Young digital citizens have significant privacy concerns due to worries about data misuse and a perceived lack of control over their personal information, worsened by frequent data breaches and unauthorized access [30], [44], [57], [58]. This highlights the need for more transparency and greater involvement of young users in how data practices are developed. Empowering young users through better data management tools can significantly reduce their discomfort and build trust in AI technologies. Although young users appreciate the benefits of AI, such as personalized services and educational tools [22], [23], they are concerned about the risks to their personal data [8], [53]. A balanced approach is required to protect privacy while maintaining the benefits of AI. Transparent data practices and policies that clearly communicate how personal information is safeguarded can help mitigate these concerns and foster a more trusting relationship with AI technologies. Additionally, young users' willingness to share data is closely linked to perceived benefits and security assurances and parental guidance also plays a role in their data-sharing practices. Educational initiatives that clarify data practices and emphasize the importance of maintaining privacy can positively impact young users' attitudes toward data sharing.

### B. Trust and Ethical Considerations

For young digital citizens, establishing trust and addressing ethical concerns is essential, given their heightened awareness of the need for transparency, reliability, and ethical conduct in digital interactions. For these users, trust in AI systems is built on clear and honest communication about data collection, usage, and protection. Transparency in data practices helps young users understand and manage their personal information effectively. Reliability is also essential, as consistent and predictable behavior from AI systems strengthens users' confidence in their functionality and safety. Ethical behavior, including adherence to privacy standards and responsible data management, aligns with the values of young users. Additionally, user-friendly interfaces are crucial for enhancing trust by simplifying privacy management and clarifying data practices.

### C. Contextual Variations in Transparency Expectations

Different environments and applications can shape how young users perceive the need for transparency in AI systems. For instance, in educational settings, students may expect detailed information about how their data is used to enhance learning tools and protect their privacy. In healthcare, transparency is crucial for ensuring that personal health data is handled responsibly, and that AI-driven diagnostics and treatments are explained clearly. Conversely, in social media platforms, where engagement is often more casual, transparency might focus on how algorithms affect content visibility and user interactions, and the expectations may be less stringent due to the informal nature of these platforms. Additionally, cultural and regional differences can affect transparency expectations, with varying levels of scrutiny and trust depending on local norms and regulations. Understanding these contextual variations is essential for designing AI systems that meet diverse transparency expectations and build trust among young digital citizens.

### D. Strategies for Enhancing User Control

Enhancing user control over personal data is crucial for addressing privacy concerns among young digital citizens. Key strategies include designing intuitive interfaces that make managing privacy settings straightforward and user-friendly. Providing users with the ability to view, update, and delete their personal data fosters a greater sense of security and control. Additionally, transparent and straightforward consent processes are necessary to build trust and ensure users feel they have control over their data. Educational initiatives that focus on data privacy and control strategies are also important, as they help young users understand their rights and how to exercise them effectively. These initiatives foster a sense of autonomy and empowerment, which is essential for navigating the digital landscape confidently.

### E. Implications for Stakeholders

The findings from this review offer valuable insights for policymakers, educators, and developers of AI applications, emphasizing the necessity for user-centered and ethically responsible AI systems that meet the privacy expectations of young users. There is a clear demand for government regulation to address AI technology, data collection, and user privacy, reflecting a widespread recognition of the need for such measures. Regulations should not be approached solely from a technical perspective but must consider diverse stakeholders to effectively safeguard user privacy rights. Enhancing digital literacy, particularly in AI, is crucial for preparing individuals, especially young digital citizens, for future risks associated with AI systems. Educational programs should integrate AI literacy and cybersecurity to equip young people with essential knowledge about privacy practices. Policymakers should develop robust and ethical frameworks to protect privacy, considering the interconnected nature of AI systems across platforms, ensuring comprehensive privacy and data protection. The study also highlights the importance of incorporating user feedback, prioritizing transparency, and upholding ethical standards in AI development. These practical and theoretical implications inform the creation of privacy policies and educational programs that address the unique concerns of young digital citizens and contribute to understanding how privacy perceptions and behaviors are influenced by sociocultural factors and AI integration into daily life, providing insights for future research and adaptive, context-specific ethical frameworks.

### F. Limitations

This study has some limitations. It relied solely on academic sources, which may not capture the most recent technological developments or practical insights from industry and non-academic stakeholders. The focus on English-language articles limits the generalizability of findings to non-English-speaking regions. Furthermore, the rapid pace of AI development means that the literature may not fully cover the latest platforms and tools, such as advanced conversational models, or cutting-edge generative AI systems. These emerging technologies introduce new privacy considerations, so our study reflects the current state of research, which is continuously evolving with technological advancements and should be viewed as a snapshot in time rather than a definitive account of the field.

## V. Conclusion

This review examines young digital citizens' perceptions, concerns, and expectations regarding privacy in AI. Key issues include a perceived lack of control over personal data, fears of misuse, and frequent data breaches. While young users value AI's benefits, such as personalized experiences and educational tools, they find it difficult to balance these advantages with privacy concerns. Though generally open to sharing data, they expect transparency in how AI processes it. From 2,000 initial papers, the review narrowed down to 108, revealing that trust in AI depends on transparency, reliability, and ethical practices. Clear communication, user-friendly interfaces, strong data protection, parental guidance, and educational initiatives are essential for fostering trust. The review also identifies gaps in research, such as the need for longitudinal studies, cultural comparisons, and practical applications of privacy-enhancing technologies. Collaboration between educators, policymakers, and AI developers is vital to integrate user feedback and uphold ethical standards, driving the development of privacy-preserving AI. Our future work will evaluate young users' engagement with AI, confidence in data sharing, and trust in company practices, while gathering insights from all stakeholders to develop guidelines for transparent, ethical, and user-centric AI systems.

## Acknowledgement

This project has been funded by the Office of the Privacy Commissioner of Canada (OPC); the views expressed herein are those of the authors and do not necessarily reflect those of the OPC.